\documentclass[prb,twocolumn,floatfix,showpacs]{revtex4-1}
\usepackage{graphicx,color}
\usepackage{amsmath,amssymb,bm}

\usepackage[export]{adjustbox}

\usepackage{physics}
\usepackage{siunitx}
\usepackage{csquotes}
\newcommand{\tn}{\textnormal}

\definecolor{darkred}{rgb}{0.90,0,0}
\definecolor{darkgreen}{rgb}{0,0.60,.2}
\definecolor{darkblue}{rgb}{0,0,1}
\definecolor{grey}{cmyk}{0,0,0,0.25}
\definecolor{orange}{cmyk}{0,0.6,0.8,0}

\newcommand{\refeq}[1]{Eq.~(\ref{#1})}

\usepackage{times}

\begin{document}

\title{Entanglement and spectra in topological many-body localized phases}


\author{K.~S.~C.~Decker,$^1$  D.~M.~Kennes,$^{2,3}$ J.~Eisert,$^4$ and C.~Karrasch$^1$}

\affiliation{$^1$Technische Universit\"at Braunschweig, Institut f\"ur Mathematische Physik, Mendelssohnstraße 3, 38106 Braunschweig, Germany}
\affiliation{$^2$Institut f\"ur Theorie der Statistischen Physik, RWTH Aachen University and JARA-Fundamentals of Future Information Technology, 52056 Aachen, Germany}
\affiliation{$^3$Max Planck Institute for the Structure and Dynamics of Matter, Center for Free Electron Laser Science, 22761 Hamburg, Germany}
\affiliation{$^4$Dahlem Center for Complex Quantum Systems and Fachbereich Physik, Freie Universit\"at Berlin, 14195 Berlin, Germany}

\begin{abstract}
Many-body localized systems in which interactions and disorder come together defy the expectations of quantum statistical mechanics: In contrast to ergodic systems, they do not thermalize when undergoing non-equilibrium dynamics. What is less clear, however, is how topological features interplay with many-body localized phases as well as the nature of the transition between a topological and a trivial state within the latter. In this work, we numerically address these questions, using a combination of extensive tensor network calculations, specifically DMRG-X, as well as exact diagonalization, leading to a comprehensive characterization of Hamiltonian spectra and eigenstate entanglement properties.

\end{abstract}

\maketitle



\section{Introduction}

The paradigm of many-body localization (MBL) uplifts Anderson localization to a regime in which genuine interactions matter. \cite{Basko,1403.7837} It manifestly breaks ergodicity and expectations from quantum statistical mechanics:\cite{HuseReview} once pushed out of equilibrium, \cite{1408.5148,PolkovnikovReview,christian_review} such quantum many-body systems will equilibrate, but retain too much memory of the initial conditions to fully thermalize. Indeed, the phenomenon of many-body localization is multi-faceted, giving rise to a plethora of phenomena that seem to have little in common at first sight. It is accompanied by an extensive number of quasi-local constants of motion, \cite{huse2014phenomenology} by eigenstates in the bulk that generically exhibit entanglement area laws, \cite{Bauer,1409.1252} and by a peculiar logarithmic dynamic growth of entanglement entropies. \cite{Prosen_localisation,Pollmann_unbounded} Reflecting this rich phenomenology, it is no surprise that the transition between an ergodic and localized regime has moved into the focus of attention. \cite{PalHuse,PhysRevLett.113.107204,PhysRevX.5.031032,Laflorencie2015,PhysRevX.5.031033}

What is less clear, at the same time, is how types of order come to play in this state of affairs. 
It has been shown that excited eigenstates can exhibit signatures of topological order,
\cite{1304.1158,PhysRevB.89.144201,PhysRevB.89.144201,Bahri} yet the precise mechanism let alone the transition to the ergodic regime are not fully understood. This is partially due to a lack of methods to address this question. Tensor network methods \cite{Orus-AnnPhys-2014,VerstraeteBig,EisertTensors} have been extended to be able to address properties of highly excited states, \cite{Kennes15,Lim16,Yu17} prominently the DMRG-X method, \cite{Khemani} generalizing the density matrix renormalization group (DMRG) method \cite{DMRGWhite92} to capture highly excited states that feature an entanglement area-law. 

\begin{figure}[t!]
\includegraphics[width=0.95\linewidth,clip]{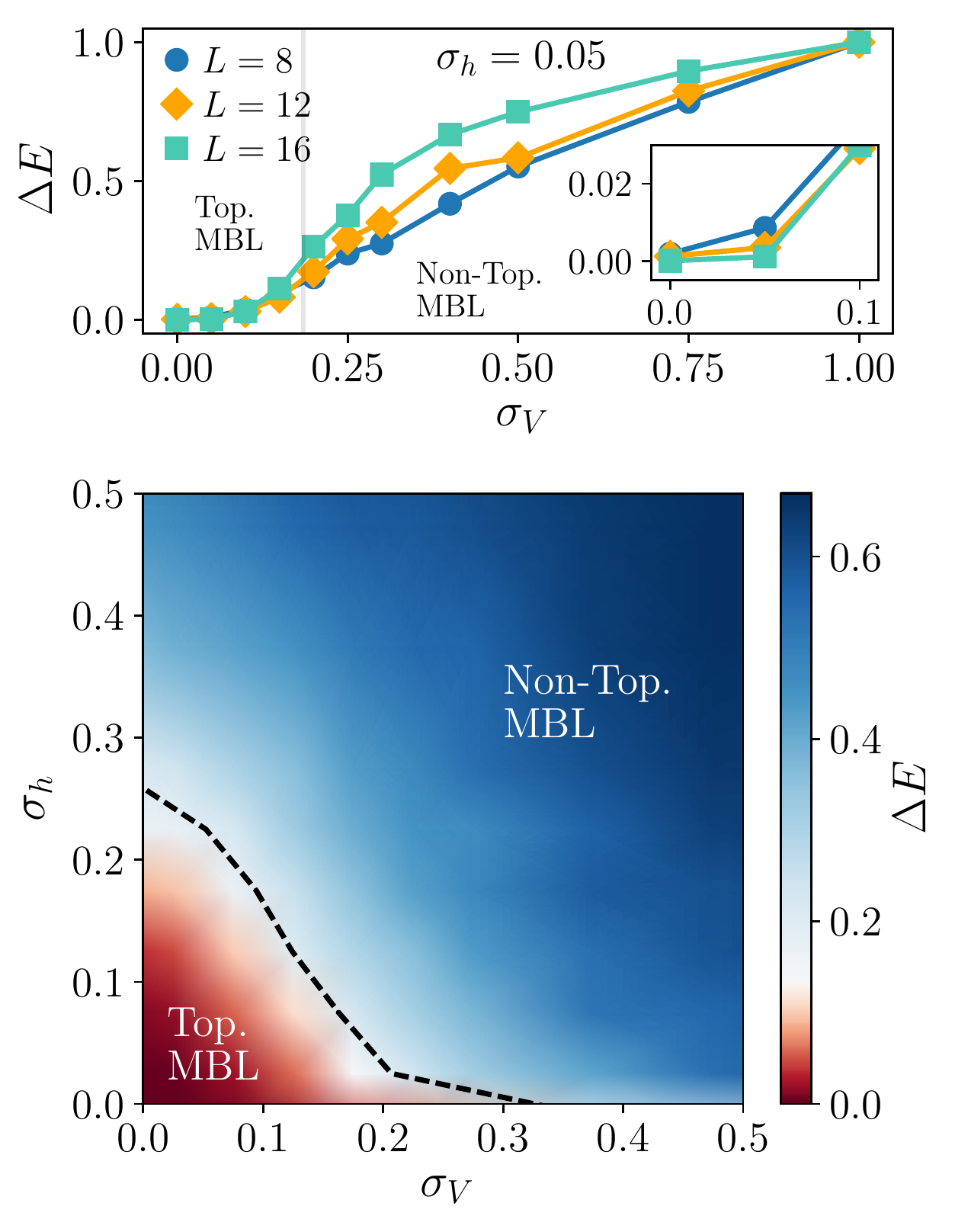}
\caption{\textit{Top panel:} Normalized distance $\Delta E$ between the highest and lowest energies of four consecutive mid-spectrum (high-energy) eigenstates as a function of $\sigma_V$ for fixed $\sigma_h=0.05$ and various system sizes for OBC. The window from which the eigenstates are taken ranges from $0.4$ to $0.6$ in terms of the reduced energy $\epsilon$ with $\Delta E$ being normalized against its maximum at $\sigma_h = 0.05$ and $\sigma_V = 1.0$. The topological phase has a four-fold degenerate spectrum (associated with the edge excitations) and is characterized by $\Delta E\to 0$ in the thermodynamic limit. The eigenstates are non-degenerate in the trivial phase ($\Delta E>0$). \textit{Bottom panel:} High-energy (mid-spectrum) phase diagram in the $\sigma_h-\sigma_V$ plane for $L=14$. This time, $\Delta E$ is normalized against $\sigma_h = 1.0$ and $\sigma_V = 1.0$. The crossover between the topological and trivial phase (dashed line) is defined as $\Delta E=0.2$. The system remains many-body localized for any of these parameters (see Fig.\ \ref{fig2}). }
\label{fig1}
\end{figure}

In contrast, addressing the question how MBL and topological properties compete or co-exist has only recently moved into the center of attention. Naively, one would expect topological order to be absent in a regime in which disorder dominates driving the system into a MBL state. However, recently a model was proposed exemplifying that this is not always the case. \cite{Bahri,PhysRevB.89.144201} We elaborate on this question using a combination of exact diagonalization of up to $L=16$ sites as well as extensive tensor network, specifically DMRG-X, approaches to determine the topological properties in the presence of MBL. By studying typical hallmarks of MBL physics as well as topology, such as the energy level statics, fluctuations of local observables and entropy, we determine the phase diagram of the model proposed in Refs.~\onlinecite{Bahri,yao15}, with variants discussed in Refs.\ \onlinecite{PhysRevB.89.144201,Tarantino}. We do so in dependence of its genuine interaction parameters, see Fig.~\ref{fig1}.

\section{Model, Simple Limits}

To be concrete, we focus on a specific model Hamiltonian, yet one that clearly shows the signatures at the heart of our argument.
The system is governed by \cite{Bahri,yao15}
\begin{equation}\label{eq:h}
 H = \sum_i \left ( \lambda_i \sigma_{i-1}^z\sigma_i^x\sigma_{i+1}^z + h_i\sigma_i^x + V_i\sigma_i^x\sigma_{i+1}^x \right) ,
\end{equation}
where $\sigma_i^{x,y,z}$ denote Pauli matrices supported on site $i$. Unless mentioned otherwise, we will work with a system of size $L$ and either open (OBC) or periodic (PBC) boundary conditions. The real pre-factors $\lambda_i$, $h_i$, and $V_i$ are random variables drawn from a Gaussian distribution with a standard deviation of $\sigma_{\lambda,h,V}$, and all data is averaged over 100-500 disorder realizations. We choose $\sigma_\lambda=1$ for the rest of this work. Note that for $V_i=0$, the system can be mapped to non-interacting fermions via a Jordan-Wigner transformation.

For $h_i=V_i=0$, the Hamiltonian of \refeq{eq:h} takes the form of a sum of mutually-commuting operators, $H_0=\sum_i\lambda_iO_i$ with $O_i=\sigma_{i-1}^z\sigma_i^x\sigma_{i+1}^z$ and $[O_i,O_j]=0$. \cite{Bahri} It can thus be treated analytically. A system with open boundaries features free spin-1/2 edge excitations generated by the Pauli operators $\Sigma^x_L=\sigma_1^x\sigma_2^z$, $\Sigma^y_L=\sigma_1^y\sigma_2^z$, $\Sigma^z_L=\sigma_1^z$ (and similarly at the right end of the chain), which commute with $H_0$. Each eigenvalue of $H_0$ is thus four-fold degenerate, and the system is in a topological phase at arbitrary energies.

One can show that each eigenstate of $H_0$ can be expressed as a matrix-product state
(MPS) with a 
bond dimension of $\chi=2$ and hence features an entanglement area law, \cite{AreaReview} 
which can be seen as a signature of localization. To this end, we rewrite $H_0 = e^{ih}\tilde H_0 e^{-ih^\dagger}$ with $h=\sum_i \sigma^x_i\sigma^x_{i+1}$ and $\tilde H_0=\sum_i\lambda_i\sigma^z_i$. The eigenstates of $\tilde H_0$ are product states. The operator $e^{ih}$ simply acts 
as a product of 
mutually commuting two-site quantum gates and transforms each product eigenstate state into matrix product states with a bond dimension of two upon conjugation: For every product state vector $|P\rangle$, $e^{ih}|P\rangle$ is such a matrix product state vector. More formally speaking, $H_0$ is an example of a class of Hamiltonians that feature exact matrix product eigenstates by construction. If $\tilde H_0$ is a $1$-local Hamiltonian with product eigenstates and $h$ a $k$-local Hamiltonian consisting of mutually commuting terms, then $H_0 = e^{ih}\tilde H_0 e^{-ih^\dagger}$ is a $2k-1$ local Hamiltonian. Since the eigenstates are obtained from products under conjugation, each eigenstate is a matrix product state of bond dimension at most $2^{(2k-1)/2}$, and hence strictly localized, satisfying an exact entanglement area law for every Renyi entropy. \cite{AreaReview}

In the converse limit where $h_i$ and $V_i$ are large, the Hamiltonian takes the form of a 
classical Ising model whose eigenstates are product states in the $\sigma^x$-basis; they are thus trivially localized but do not feature topological properties. This suggests that the system may be 
localized for any values of $\sigma_{h,V}$ but that a transition between a topological MBL phase and a trivial MBL phase occurs when $\sigma_{h,V}$ are increased. We will now confirm this scenario explicitly using a combination of exact diagonalization and DMRG-X numerics.

\section{Phase diagram from exact diagonalization}
\subsection{Topology: Spectrum degeneracy}

We first show that the topological properties of highly-excited states survive for $h_i\neq0$, $V_i\neq0$. To this end, \refeq{eq:h} is solved for OBC by exact diagonalization. The topological phase features a spin-$1/2$ degree of freedom at each edge and is thus characterized by a four-fold degeneracy of the spectrum in the thermodynamic limit (finite systems feature exponential corrections). The trivial phase exhibits a non-degenerate spectrum.

In the top panel of Fig.\ \ref{fig1}, we plot the normalized distance between the four consecutive eigenvalues $\Delta E$, which serves as a measure for the degeneracy of the spectrum, as a function of $\sigma_V$ for fixed $\sigma_h=0.05$ and various system sizes $L$. Here, $\Delta E$ is normalized against its maximum (at $\sigma_h=0.05$ and $\sigma_V=1.0$), with the phase boundary being defined at $\Delta E=0.2$ in this normalization. The disorder sampling has been carried out such that the error is smaller than the symbol size, and we limited ourselves to mid-spectrum (high-energy) states in a window $[0.4,0.6]$ in terms of the reduced energy $\epsilon=(E-E_{\rm min})/(E_{\rm max}-E_{\rm min})$, with $E_{\rm min}$ and $E_{\rm max}$ the minimum and maximum energy of the spectrum, respectively. For small (large) $\sigma_V$, $\Delta E$ decreases (increases) with $L$. The data of Fig.\ \ref{fig1} thus indicates that a transition between a high-energy topological phase and a trivial phase occurs around $\sigma_V\sim 0.2$. In the bottom panel of Fig.\ \ref{fig1}, we show the degeneracy $\Delta E$ for fixed $L=14$ in the $\sigma_h-\sigma_V$ plane. The phase boundary (dashed line) has been defined via $\Delta E=0.2$, with $\Delta E$ again being  normalized against its maximum (which occurs at $\Delta E$ at $\sigma_V=1.0$ and $\sigma_h=1.0$).

\begin{figure}[t]
\includegraphics[width=0.95\linewidth,clip]{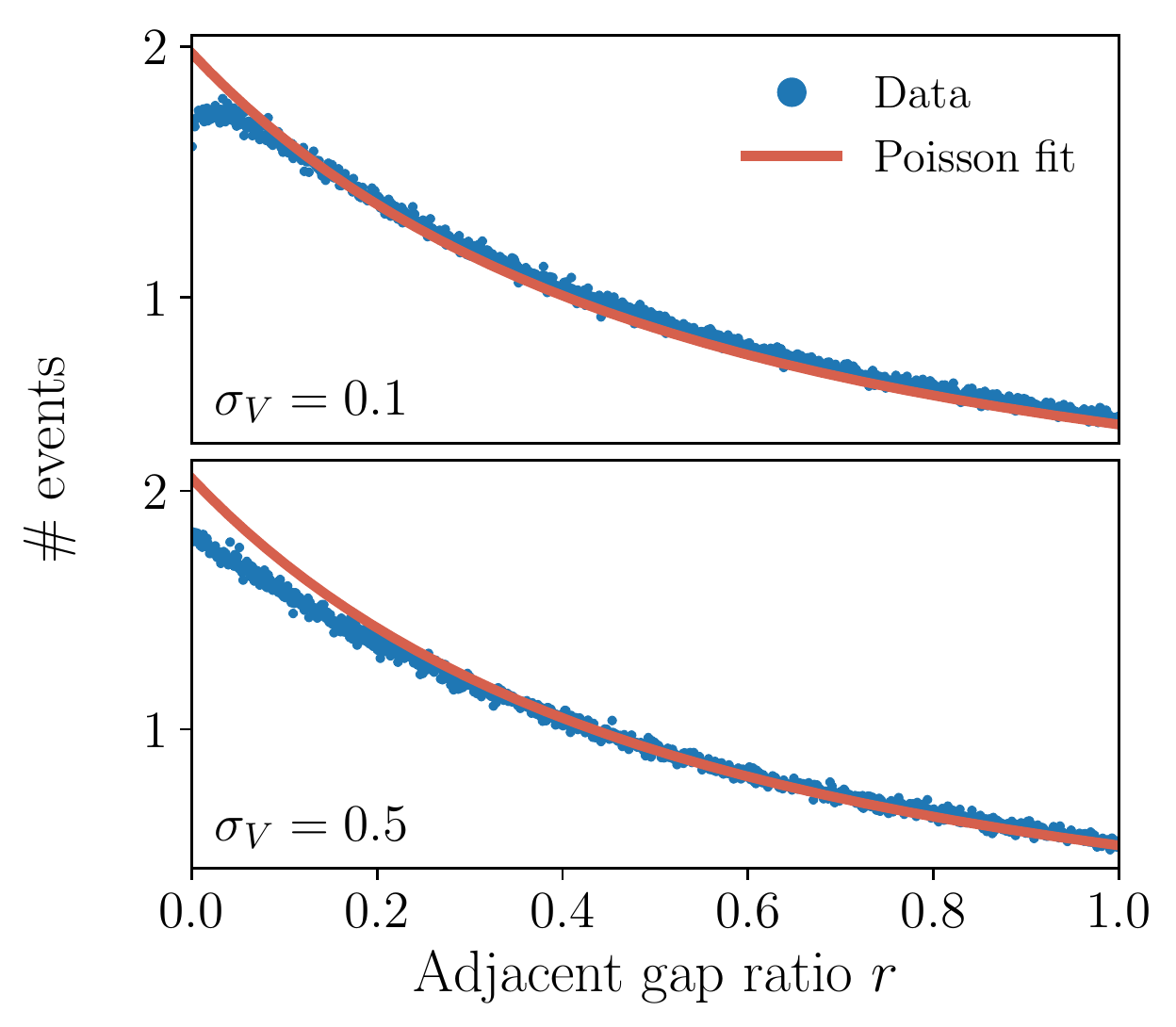}
\caption{Distribution of the adjacent gap ratio in mid-spectrum eigenstates for fixed $\sigma_h=0.05$, various $\sigma_V$, and $L=16$ for PBC. The AGR has a Poissonian form, which is a hallmark of localization. The eigenstates are all from the interval $[0.4,0.6]$ in terms of the reduced energy $\epsilon$. }
\label{fig2}
\end{figure}

\begin{figure}[t]
\includegraphics[width=0.95\linewidth,clip]{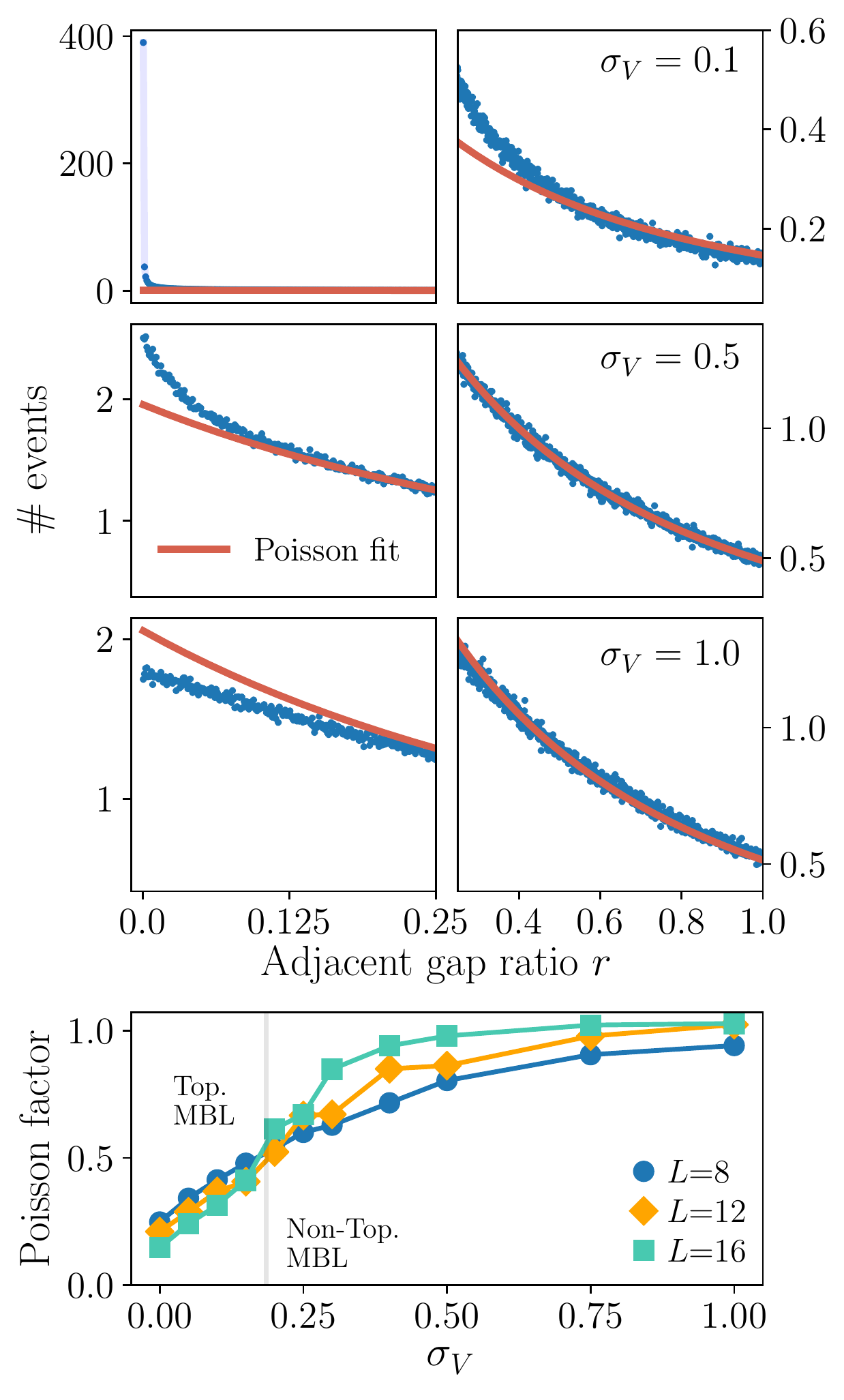}
\caption{\textit{Top panels:} The same as in Fig.\ \ref{fig2}, but for OBC. The topological phase features a sharp peak at $r=0$ due to the quadrupel of almost-degenerate eigenstates, the tail is always Poissonian. \textit{Bottom panel:} Relative weight of the Poissonian tail for fixed $\sigma_h=0.05$ and various $L$ for a system with OBC.}
\label{fig2old}
\end{figure}

\subsection{Many-body localization: Adjacent gap ratio}

As a next step, we provide strong evidence that the system remains localized for arbitrary values of $\sigma_h$ and $\sigma_V$. We first focus on a system with PBC in order to remove the spectral degeneracy associated with the edge degrees of freedom. In Fig.\ \ref{fig2}, we show the distribution of the adjacent gap ratio \cite{Oganesyan} (AGR) $r$ for the mid-spectrum states $\epsilon\in [0.4,0.6]$ for fixed $\sigma_h=0.05$, various $\sigma_V=0.1,0.5$, and $L=14$. The distribution always takes a Poissonian form, which is a hallmark of localization. \cite{Oganesyan}

For a system with OBC, the AGR features a sharp peak at $r=0$ in the topological phase (see Fig.\ \ref{fig2old}), which signals the extensive number of quadruples of (close to degenerate) eigenstates associated with the edge excitations; the tail of the distribution is always Poissonian. The relative weight of the Poissonian tail as a function of $\sigma_V$ for fixed $\sigma_h=0.05$ is shown in the bottom panel of Fig.\ \ref{fig2old}. For small (large) $\sigma_V$, this weight decreases (increases) with the system size, implying that more (less) weight is shifted into the peak at $r=0$. This provides further evidence that a transition between a topological and a trivial phase takes place around $\sigma_V\sim0.2$. 

\begin{figure}[t]
\includegraphics[width=0.95\linewidth,clip]{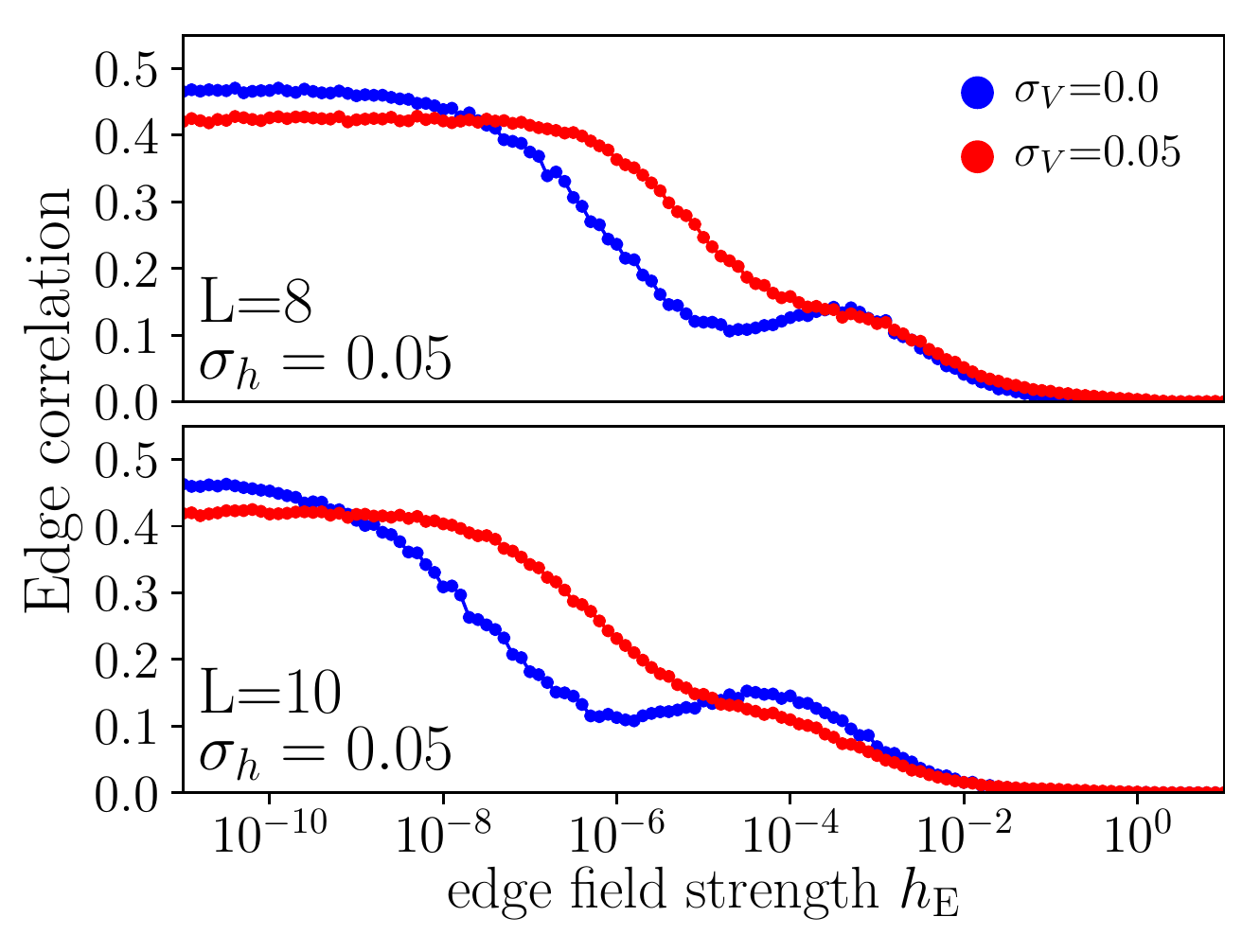}
\caption{Coupling between the spin-$1/2$ degrees of freedom associated with the two edges of a finite system with OBC. The spins can be decoupled by switching on a small magnetic field $h_\tn{E}$ at the ends of the system.}
\label{fig:edgefield}
\end{figure}

\section{Entanglement properties and bi-partite fluctuations from exact diagonalization}

Next, we turn to discussing physical properties of the system such as the entanglement entropy or bi-partite fluctuations. We first present ED data; our aim is to eventually use the DMRG-X to access large system sizes of up to $L\sim50$ sites. We exclusively focus on systems with OBC from now on.

In the topological phase with generic $\sigma_h\neq0$, $\sigma_V\neq0$, the spectrum for finite $L$ consists of almost-degenerate quadruplets associated with the spin-$1/2$ degrees of freedom at each edge. It turns out that the left and right edge spins are coupled in the eigenstates of a finite systems. Such a coupling is not stable towards small perturbations. In order to extract the generic behavior of physical quantities, one can pursue two different strategies: One can either compute bulk properties which are not affected by a coupling between the edges, or one can remove this coupling by switching on a small edge field $\pm h_\textnormal{E}\Sigma_{L,R}$ which splits up the quadruplets in the thermodynamic limit. The latter is illustrated in Fig.\ \ref{fig:edgefield}, which shows the two-point correlation function $\frac{1}{4}\sum_{i,j\in x,z}|\langle\Sigma^i_L\Sigma^j_R\rangle-\langle\Sigma^i_L\rangle\langle\Sigma^j_R\rangle|$ as a function of the edge field.

We first investigate bulk properties which are not affected by a potential coupling between the edges. The standard bi-partite entanglement entropy is certainly no such quantity, and it is generally not even well-defined in a system with a degenerate spectrum (i.e., in the thermodynamic limit) as it intertwines classical and genuine quantum correlations. From the perspective of entanglement theory, the entanglement entropy is a valid entanglement measure \cite{Horodecki_entanglement,InHouseReview} for pure but not for mixed quantum states. For this reason, we resort to the logarithmic entanglement negativity. \cite{Volume,EisertPlenioNeg} Both the negativity and its logarithmic counterpart are faithful entanglement measures also for mixed quantum states. \cite{PhD,VidalNegativity,PlenioNegativity}
It is defined as
\begin{equation}
    E_N(\rho) = \log_2\|\rho^\Gamma\|_1
\end{equation}
for a (pure of mixed) density operator $\rho$, where
$\rho^\Gamma$ denotes the partial transpose of $\rho$ in any basis of a distinguished subsystem. We stress that the use of the entanglement negativity instead of the entanglement entropy is required in any quantum many-body setting in which degeneracies in sub-spaces are becoming exponentially small in the system size.

\begin{figure}[t]
\includegraphics[width=0.95\linewidth,clip]{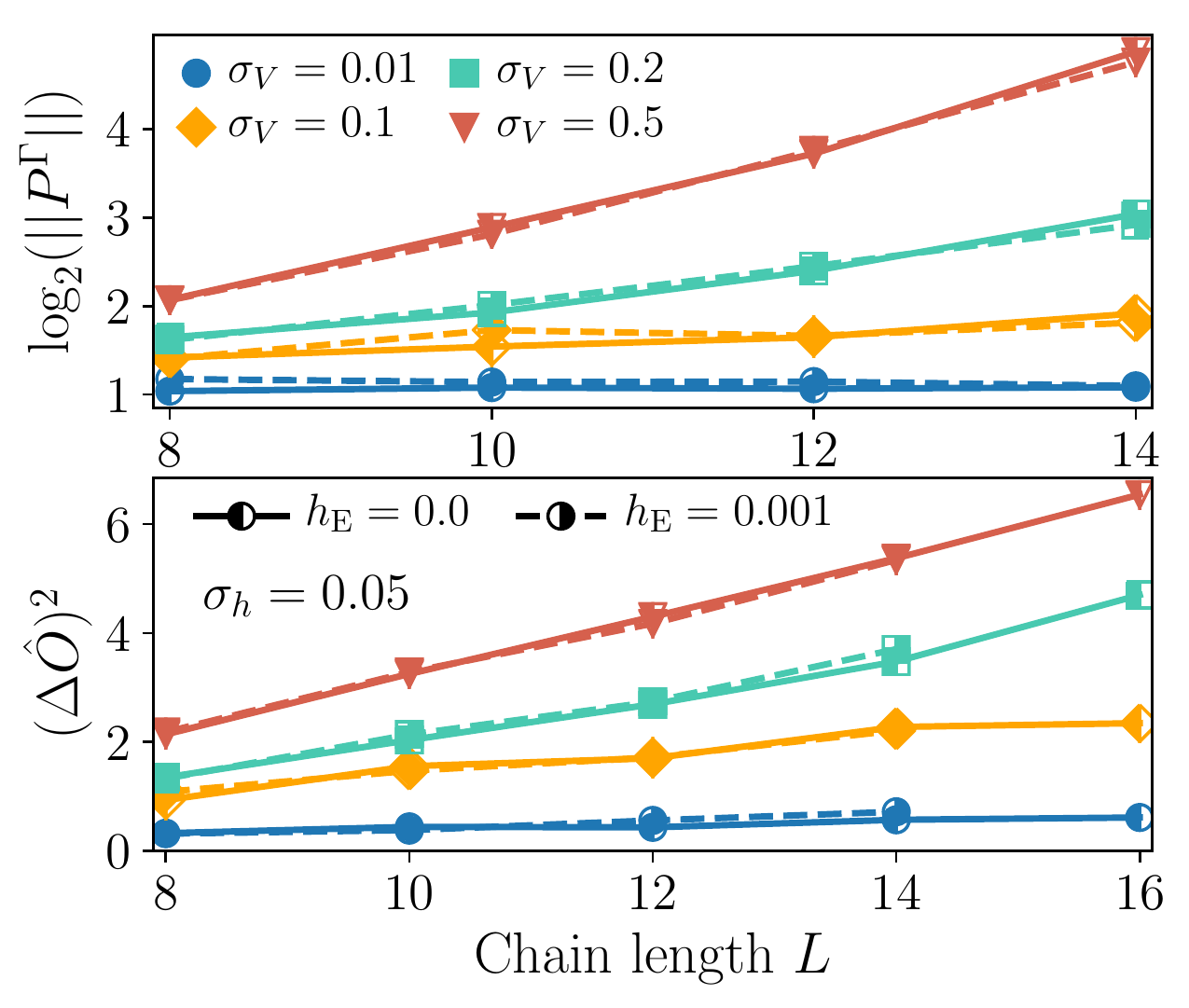}
\caption{Scaling of the entanglement negativity (upper panel) and the three-site spin fluctuations (lower panel) with the system size for fixed $\sigma_h=0.05$ and various $\sigma_V$ both with and without an additional edge field which removes the coupling between left and right spin-$1/2$ degrees of freedom. The localization length becomes larger than $L=16$ for $\sigma_V\sim 0.15$ (the AGR indicates that the system is localized for any $\sigma_V$, see Fig.\ \ref{fig2}).}
\label{fig3}
\end{figure}

In addition to the entanglement negativity, we study the bi-partite fluctuations of $\{O_i\}$ (the analogue of the bi-partite spin fluctuations considered for the XXZ chain). Results for both quantities are shown in Fig.\ \ref{fig3} as a function of the system size both with and without an additional edge field, confirming that those bulk properties are indeed not affected by a potential coupling between the edges. Concomitant with many-body localization, we expected generic eigenstates to exhibit an area (volume) law \cite{AreaReview} in a localized (ergodic) phase. For small $\sigma_V$, we indeed observe an area law. At $\sigma_V\sim0.1$, both quantities saturate around $L=14$, and for even larger $\sigma_V$, we see volume-law behavior on the accessible system sizes. Since the AGR suggests that the system is still in a localized phase, this indicates that the localization length becomes larger than $L$ for these parameters, which in turn casts doubts whether or not the transition from the topological to the trivial phase, which we observe for similar parameters (see Fig.\ \ref{fig1}), is an artifact of small system sizes and that the topological phase in fact remains stable for larger values of $\sigma_h$ and $\sigma_V$ (note, however, that the system becomes a trivial insulator in the Ising limit $\sigma_h,\sigma_V\to\infty$).

\section{DMRG-X}
\subsection{General idea}

The DMRG-X method is a tensor network method that has been introduced in Ref.\ \onlinecite{Khemani} as a tool to determine the matrix product state representation of a highly-excited but localized eigenstate in a disordered system in one spatial dimension. The fact that eigenstates generically satisfy area laws for entanglement entropies in the many-body localized regime \cite{HuseReview,Bauer} renders the method applicable. The method has been developed and tested for the disordered XXZ chain governed by $H=\sum_i (\sigma^x_i \sigma^x_{i+1}+ \sigma^y_i \sigma^y_{i+1}+ \sigma^z_i\sigma^z_{i+1}+h_i\sigma^z_i)$, where $h_i$ are random fields for each $i$. The key idea is to prepare the system in a random product state in $z$ direction, which becomes an eigenstate of $H$ in the limit of large $h_i$. Thereafter, DMRG sweeps are carried out and the bond dimension $\chi$ is successiley increased, but instead of choosing the lowest-energy state during each 1 or 2-site DMRG step (as is done for a ground state calculation), one picks the state which has maximum overlap with the prior state. This accounts for the fact that localized eigenstates which have similar energy differ vastly in their spatial structure, and one can determine the matrix product state representation of an excited eigenstate with up to machine precision.

\begin{figure}[t]
\includegraphics[width=0.95\linewidth,clip]{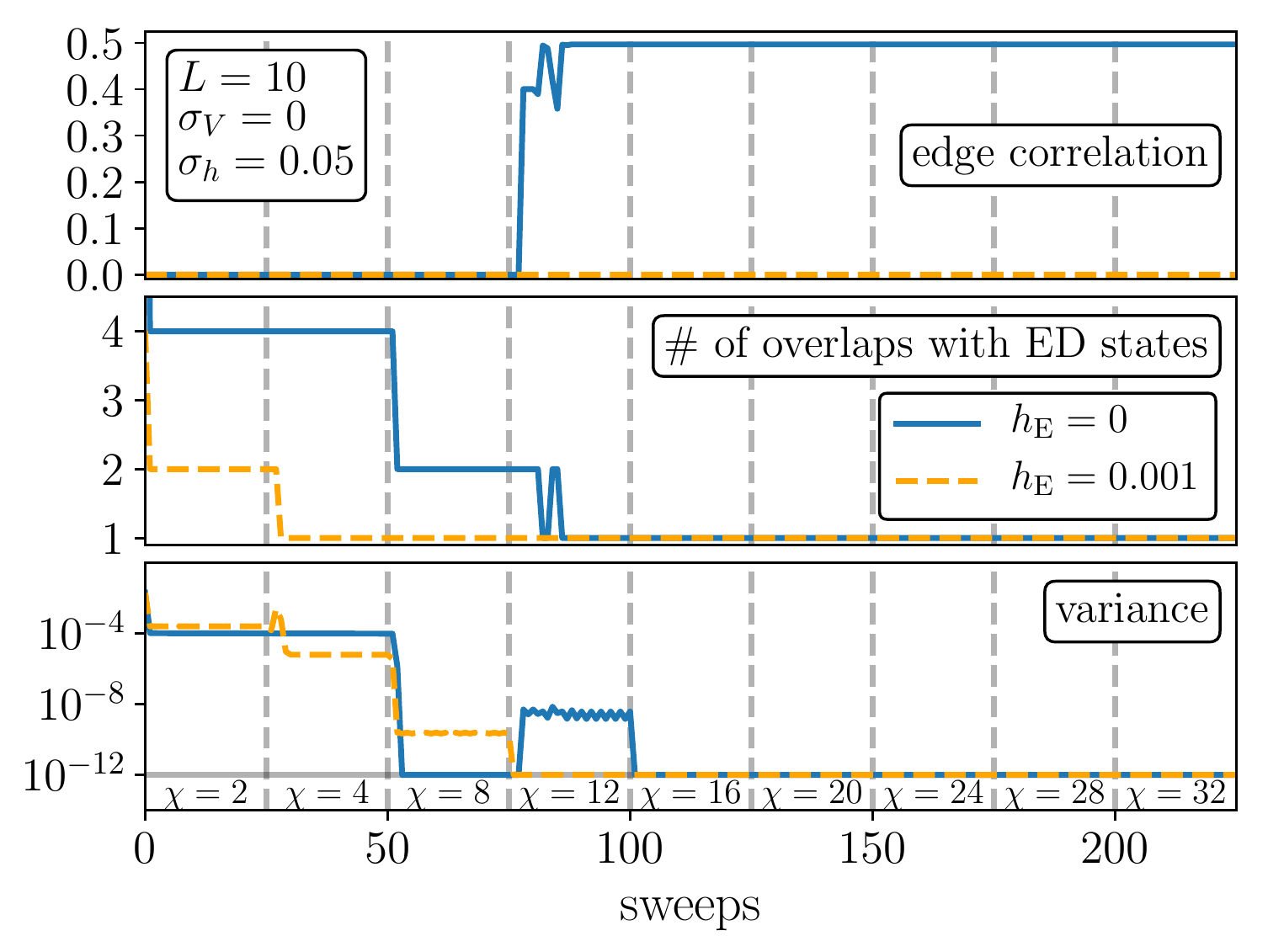}
\caption{Evolution of the two-point correlation function between the edge spins (top panel), the number of states in the ED spectrum that current DMRG-X state has a non-zero overlap with (middle panel), and the variance of the energy $\langle H^2\rangle - \langle H\rangle^2$ (bottom panel) as a function of the DMRG-X sweeps for a fixed disorder configuration. The left and right edge spins are decoupled in the initial state. In the absence of an edge field ($h_\textnormal{E}=0$), the spins are coupled as the bond dimension $\chi$ is increased during the sweeps, and one obtains an exact eigenstate of the system. The spins stay decoupled for $h_\tn{E}\neq0$. The initial state and disorder configuration are identical in both cases.}
\label{fig:sweeps1}
\end{figure}

\subsection{Application to our model}

This DMRG-X method can be generalized straightforwardly to the Hamiltonian at hand. One prepares the system in an eigenstate of $H_0$, each of which can be written as a matrix product state with a bond dimension of $\chi=2$. In practice, these states can be constructed using a simple recursive algorithm; we choose them such that the left and right edge spins are not coupled (which is possible since all eigenstates are strictly four-fold degenerate for $\sigma_h=\sigma_V=0$ even in a finite system).

We first focus on small systems and discuss whether or not the states obtained by the DMRG-X are indeed exact eigenstates. As mentioned above, the left and right edge spins are coupled in the eigenstates of a system with OBC for $\sigma_h\neq0$ and/or $\sigma_V\neq0$. One would expect that the DMRG-X, which is a matrix-product state method and thus biased towards low-entanglement states, does not couple the edges during the sweeps if they are initially uncoupled and does thus not yield exact eigenstates even for small systems. This is, however, not true (see Fig.\ \ref{fig:sweeps1}, left panel): As the bond dimension $\chi$ is increased during the sweeps, the DMRG-X state couples the left and right edges (top panel) and coincides with precisely one state from the entire ED spectrum for $\chi\geq16$ (middle panel). There is, however, one important caveat: Usually, the variance of the energy $\langle H^2\rangle - \langle H\rangle^2$ is used to gauge convergence. In our case, however, this variance drops to machine precision before an exact eigenstate has been obtained (see Fig.\ \ref{fig:sweeps1}, bottom panel, $\chi=8$). Only if the DMRG-X procedure is continued and the bond dimension is increased, one eventually converges to an exact eigenstate.

Fig.\ \ref{fig:sweeps1} also shows the same comparison between the DMRG-X and ED data in the presence of a finite $h_\tn{E}$ which decouples the edges (both the initial state and the disorder configuration are the same as before). In this case, the DMRG-X yields an exact eigenstate even for small $\chi$.

\begin{figure}[t]
\includegraphics[width=0.95\linewidth,clip,valign=t]{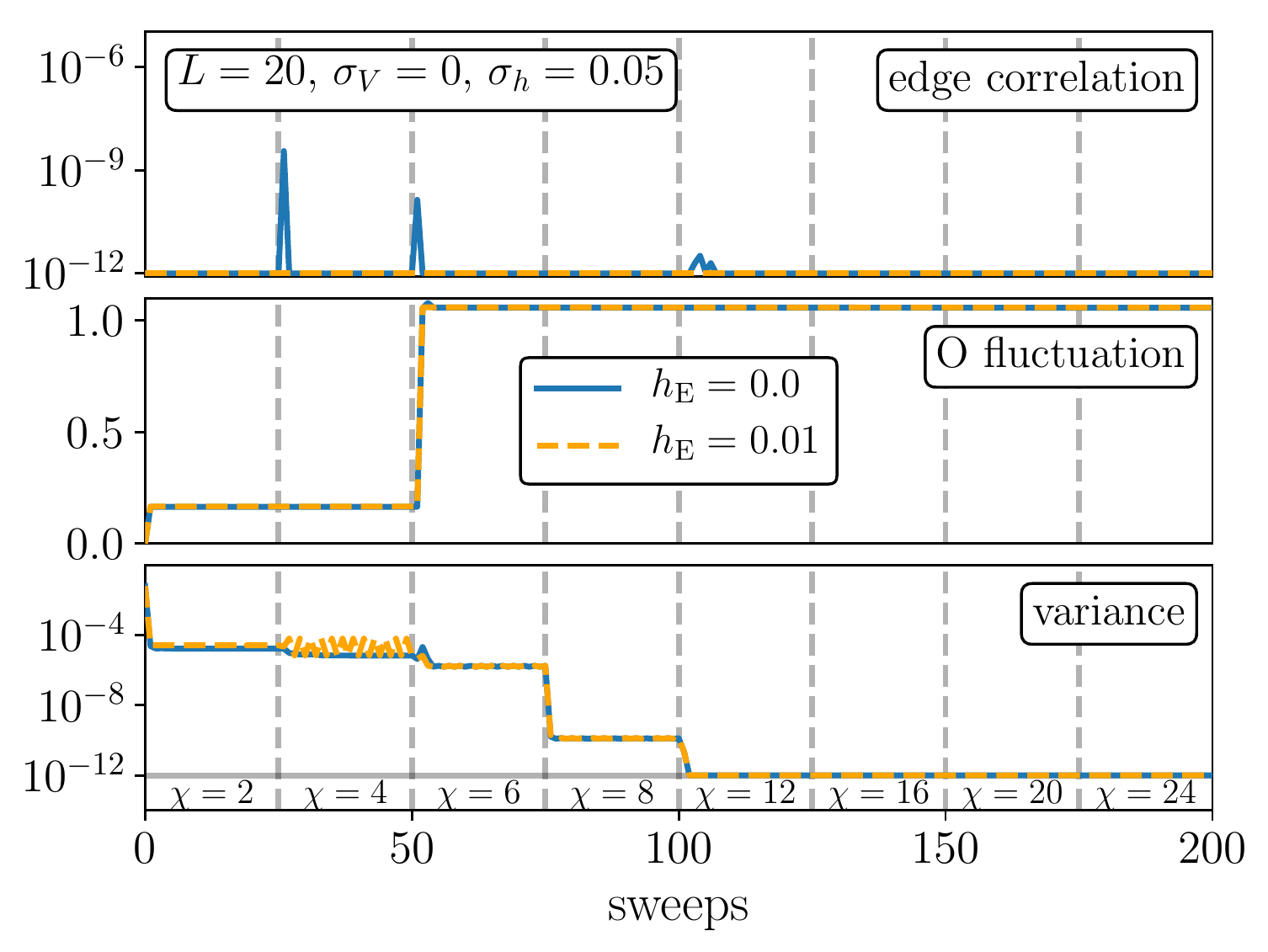}
\caption{Comparison of DMRG-X data for a larger system obtained with and without an edge field (solid blue and dashed orange lines, respectively). The initial state and the disorder configuration are the same in both cases. Both approaches yield the same state in which the edge spins are not coupled.}
\label{fig:sweeps2}
\end{figure}

Our goal is to employ the DMRG-X to study large systems which are not accessible by ED. In order to determine the nature of the states obtained at the end of the sweeps, it is instructive to compare results obtained with and without an edge field for identical initial states and disorder configurations (see Fig.\ \ref{fig:sweeps2}). It turns out that even for $h_\tn{E}=0$, the edges do not get coupled during the sweeps; thus, one does not obtain an exact eigenstate but a state that is identical to the one calculated for $h_\tn{E}\neq0$ up to a local rotation of the left and right edge spins. This is reasonable since for large $L$ the bond dimension cannot be chosen large enough to encode the entanglement between the edges, and the DMRG-X is stuck with a low-entanglement state with decoupled edges. Put differently, for large $L$ the DMRG-X automatically yields the physical state in which the edge spins are not coupled even in the absence of an edge field.

\subsection{Results}

We use the DMRG-X to compute physical quantities analogous to the ones shown in Fig.\ \ref{fig3} for larger systems. The energy variance of states obtained using the DMRG-X is machine precision.

\begin{figure}[t]
\includegraphics[width=0.93\linewidth,clip]{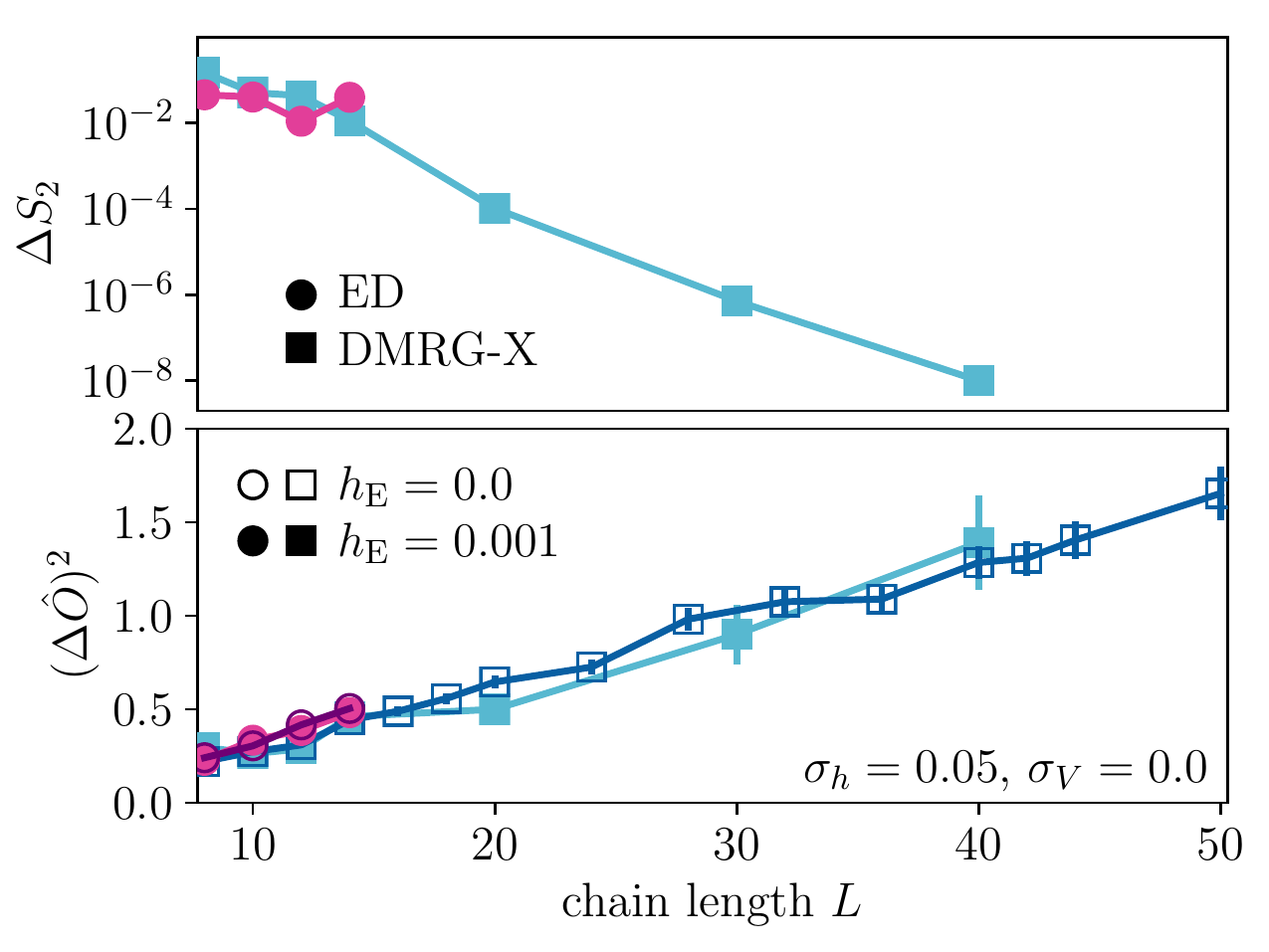}
\caption{Comparison of ED as well as DMRG-X data. \textit{Top panel:} Gap of the bi-partite entanglement spectrum (obtained using a finite edge field $h_\textnormal{E}$ to decouple the edge spins); this gap decreases exponentially with $L$, which is a hallmark of the topological phase. \textit{Bottom panel:} The bi-partite spin fluctuations, which show a slow linear or logarithmic increase despite the fact that the AGR suggests the system is in a MBL phase (see the main text for details). The calculation was performed both for $h_\textnormal{E}=0$ and $h_\textnormal{E}\neq0$; the results coincides, which one expects for this bulk quantity. }
\label{fig:dmrgx}
\end{figure}

We first investigate the behavior of the gap in the half-chain entanglement spectrum as a function of the system size (note that the entanglement negativity is not a pure-state measure and is thus inaccessible). We apply a finite $h_\tn{E}\neq0$ in order to decouple the edge degrees of freedom. Results are shown in the top panel of Fig.\ \ref{fig:dmrgx}; ED data is shown for comparison at small $L$. The entanglement gap decreases exponentially with the system size -- all eigenvalues of the reduced density matrix are (almost) doubly degenerate. This provides further evidence that the system is in the topological phase for these parameters.

Secondly, we study the bi-partite fluctuations of $O_i$. Results are shown in the bottom panel of Fig.\ \ref{fig:dmrgx} and are again consistent between DMRG-X and ED for the system sizes accessible to both. As mentioned above (see also Fig.\ \ref{fig3}), this quantity is dominated by the bulk, and one expects that it is not influenced by the fact whether or not the edges are coupled. In order to demonstrate this explicitly, we show data obtained with $h_\tn{E}=0$ where the edges are coupled for small $L$ (for large $L$, the DMRG-X automatically decouples the edges; see above).

For system sizes beyond those accessible by ED, one observes a slow growth of the bi-partite $O_i$-fluctuations, which seems consistent with either logarithmic or linear growth in system size. There are several possible interpretations for this result: (i) The growth of fluctuations is linear and becomes logarithmic or constant only at larger system sizes than accessible to us; this would however mean that the localization length is very large (although the perturbations are very small) or (ii) the localization length is actually small compared to the system sizes we study and the behavior we see is truly logarithmic, which is difficult to distinguish on the scales accessible.

\section{Conclusions}  We have characterized the interplay of many-body localization physics and topology in a spin model where these phases can coexist instead of excluding each other using a combination of extensive exact diagonalization and DMRG-X calculations. We mapped out the full phase diagram including the topological-trivial transition with the many-body localized spin system under scrutiny; the topological phase was defined via a four-fold degeneracy of all eigenvalues of a system with open boundaries, and MBL was characterized using the adjacent gap ratio. Both entanglement properties and bi-partite spin fluctuations feature area laws in the MBL phase; the gap in the entanglement spectrum vanishes in the topological phase.

Using the DMRG-X, we gain access to system sizes far beyond the limitations of exact diagonalization and find a sluggish growth of bi-partite fluctuations of local observables, which is consistent with both linear or logarithmic behavior (which is difficult to distinguish even on these larger system sizes). This leaves two likely conclusions: either the behavior is linear meaning that the localization length is surprisingly large in the system studied or the behavior is logarithmic, which unambiguously can be proven only at even larger system sizes and should be subject of future investigations. It is the hope that the present work --- bringing together ideas of condensed matter and quantum information theory 
--- provides a machinery to identify phases of matter and further flesh out the interplay of disorder and topological signatures.

{\it Acknowledgments.} This work has been supported by the DFG (CRC 183, FOR 2724, EI 519/14-1, EI 519/15-1) and through the Emmy Noether program (KA 3360/2-1), the ERC (TAQ), and the Templeton Foundation. This work has also received funding from the European Union's Horizon 2020 research and innovation program under grant agreement No 817482 (PASQuanS).

\bibliography{BigReferences42.bib}

\end{document}